\documentclass[a4paper,12pt]{article}
\usepackage[bbgreekl]{mathbbol}
\usepackage{mathrsfs}
\usepackage{graphicx}
\usepackage{amsmath}
\numberwithin{figure}{section}
\usepackage{amsfonts}
\usepackage{indentfirst}
\usepackage{amsthm}
\usepackage{amssymb}
\usepackage{xcolor}
\usepackage{geometry}
\usepackage{url}
\usepackage{setspace}
\usepackage{verbatim}
\usepackage{ulem}
\usepackage{multirow}
\usepackage{booktabs}
\usepackage[page,toc,titletoc,title]{appendix}
\usepackage[noend]{algpseudocode} 
\usepackage{algorithm}
\usepackage{subfigure}
\usepackage{caption}

\numberwithin{equation}{section}

\newtheorem{definition}{Definition}[section]

\newcommand\dd{~{\rm d}}
\newcommand\DD{\mathcal{D}}

\newcommand\R{\mathbb{R}}

\newcommand\C{\mathbb{C}}

\newcommand\RL{\mathcal{R}}

\newcommand\tpsi{\tilde{\psi}}
\newcommand\veff{V_{\rm eff}}

\newcommand\hu{\hat{u}}
\newcommand\hH{\hat{H}}

\newcommand\bG{{\bf G}}
\newcommand\bzero{{\bf 0}}

\definecolor{wt}{rgb}{0.4660 0.6740 0.1880}

\title{Localization in the Incommensurate Systems: A Plane Wave Study via Effective Potentials}

\author{
Ting Wang\thanks{Corresponding.
{\it ting\_wang@lsec.cc.ac.cn},
LSEC, Institute of Computational Mathematics and Scientific/Engineering Computing, Academy of Mathematics and Systems Science, Chinese Academy of Sciences, Beijing 100190, China.
},
~ Yuzhi Zhou\thanks{
{\it zhou\_yuzhi@iapcm.ac.cn},
CAEP Software Center for High Performance Numerical Simulation, Beijing 100088, China;
Institute of Applied Physics and Computational Mathematics, Beijing 100094, China.
}
~ and ~ Aihui Zhou\thanks{
{\it azhou@lsec.cc.ac.cn},
LSEC, Institute of Computational Mathematics and Scientific/Engineering Computing, Academy of Mathematics and Systems Science, Chinese Academy of Sciences, Beijing 100190, China;
School of Mathematical Sciences, University of Chinese Academy of Sciences, Beijing 100049, China.
}
}

\date{}

\begin{document}

\maketitle

\begin{abstract}
In this paper, we apply the effective potentials in the localization landscape theory (Filoche et al., 2012, Arnold et al., 2016) to study the spectral properties of the incommensurate systems.
We uniquely develop a plane wave method for the effective potentials of the incommensurate systems and utilize that,
the localization of the electron density can be inferred from the effective potentials. Moreover, we show that the spectrum distribution can also be obtained from the effective potential version of Weyl's law.
We perform some numerical experiments on some typical incommensurate systems, showing that the effective potential provides an alternative tool for investigating the localization and spectrum distribution of the systems.
\end{abstract}


\section{Introduction}
\label{sec:introduction}

Localization of waves in the non-periodic media is a remarkable and well-known phenomenon, which has garnered significant interest due to its critical role in numerous material properties \cite{ostrovsky2006electron}.
Due to the advances in techniques nowadays, the localization of incommensurate systems has been widely observed from experiments in optical, mechanical, and low-dimensional material systems \cite{deissler2010delocalization,roati2008anderson, segev2013anderson, wang2020localization} about 60 years after the classic work of Anderson \cite{anderson1958absence}.
In \cite{deissler2010delocalization,roati2008anderson}, the localization was studied in cold atomic gases by controlling
artificially incommensurate potentials through finely tuned lasers.
Furthermore, the localization of electronic wave functions in modern low-dimensional materials such as twisted bilayer graphene can impact drastically their transport and magnetic properties \cite{ostrovsky2006electron}. These materials form incommensurate structures through a precisely controlled interlayer twist.
And many significant physical quantities of interests, like the quantum Hall effect \cite{dean2013hofstadter}, the enhanced carrier mobility \cite{kang2017moire}, and the unconventional superconductivity \cite{cao2018unconventional} 
have deep connections with the localization of the electrons in the flat bands, though they have not been fully understood yet.
Therefore, studying the localization mechanism in incommensurate systems is a crucial step to understanding the relevant unconventional phenomena and making use of them in the future.

Theoretical studies on the localization properties of disordered or incommensurate systems have raised particular attention in the physics and mathematics communities \cite{arnold2019localization,arnold2016effective,li2017mobility,modugno2009exponential}, and references therein.
Only the literature directly relevant to this paper will be listed.
In \cite{chen2021plane}, the localized-extended transitions in the one-dimensional incommensurate systems were studied by means of a scattering picture under the plane wave discretization.
Alternatively, a new mathematical technique was proposed to approximate the eigenstates. In \cite{filoche2012universal}, the landscape concept was introduced first, whose authors provided a novel sight to predict the location of the low energy eigenfunctions by the relevant Dirichlet problem. In \cite{lu2018detecting,steinerberger2017localization}, some mathematics consequences were presented. Further, in \cite{arnold2019computing,arnold2016effective}, effective potential defined as the reciprocal of landscape function was proposed. It contains abundant information on localized eigenpairs, 
providing a wealth of insights into their localization properties with the advantage of reduced computational cost. Recently, considerable applications of this theory have been presented \cite{altmann2019localized, arnold2022landscape, chenn2022approximating, filoche2017localization,pelletier2022spectral,razo2022low}.
In \cite{filoche2017localization,pelletier2022spectral}, the localization properties in systems gases of ultracold atoms and disordered semiconductor alloys were addressed by these prediction methods.
In \cite{arnold2022landscape,razo2022low}, the localization was studied for tight-binding Hamiltonians in two-dimensional materials by means of the localization landscape theory. Yet, to our best knowledge, this framework has not been applied to the incommensurate systems.

On the other hand, spectral properties play a central role in determining the physical behaviors of incommensurate systems. 
Through the evaluation of the spectral distribution, we can in principle derive the system's conductivity, specific heat, magnetism, and superconductivity, among other relevant properties. The density of states is a powerful tool in this respect, as it allows us to compute the spectral distribution by quantifying the number of accessible energy states per unit of energy in a more or less mean-field context, and as an intermediate quantity to further calculate other physical properties.
The integrated density of states, essentially an accumulation of the density of states, offers a broader perspective by accounting for the number of states with energies less than a certain threshold.
A crucial aspect of this process is the application of Weyl's law. This fundamental mathematical and physical principle offers an asymptotic estimation of the integrated density of states, yielding precise results, particularly for large systems and high-energy scenarios \cite{zworski2022semiclassical}. In \cite{arnold2019computing,arnold2016effective}, an approximation of the integrated density of states represented by the effective potential, was constructed based on a variant of Weyl's law. This was pursued due to the limitations of the standard approximation.

In this work, we study the localization in the incommensurate system by the plane wave method with the aid of the effective potential. 
Unlike the existing literature that addresses the problem within a bounded domain equipped with a specific boundary condition, we undertake a more comprehensive approach. We directly discretize the Schr\"{o}dinger operator with incommensurate potentials across the entire real space and offer the formulation of the effective potential.
From the effective potential, we are able to extract the spectral properties of the systems within the real space domain of our interest. 
Specifically, the positions of the minima for the effective potential allow us to infer the localization positions of the electron density that served as an alternative more effective physical observable for studying localization in such extended systems. 
Moreover, we provide a means to predict the spectral distribution of incommensurate systems based on the effective potential. 
The 1D and 2D numerical experiments give examples of how to capture the localization information of the electron density and spectral distribution by the effective potential efficiently.

{\bf Outline.}
The rest of this paper is organized as follows. In Section \ref{sec:Incommeig}, the incommensurate systems, the corresponding Schr\"{o}dinger operator and 
other associated observables are briefly introduced.
In Section \ref{sec:effectiveP}, the localization landscape theory and effective potential for the incommensurate systems are presented. In Section \ref{sec:pw}, numerical schemes for partial differential equations related to the incommensurate eigenvalue problem are listed. Furthermore, the process of studying localization under the context of plane wave discretization is explored. 
In Section \ref{sec:comput}, some numerical experiments are performed to show the procedure of predicting electron density and spectral distribution from the effective potential. 
In Section \ref{sec:conclusion}, some conclusions are drawn.

\section{Localization in quantum incommensurate systems}
\label{sec:Incommeig}

Our study focuses on two $d$-dimensional ($d=1,2$) periodic systems, arranged in parallel along the $(d+1)$th dimension. For the sake of simplicity, we neglect the $(d+1)$th dimension and the distance between the two layers. Notably, the theoretical and algorithmic frameworks elaborated upon in this paper are readily generalizable to incommensurate systems with more than two layers and models that involve the $(d+1)$th dimension.

A periodic system with $d$-dimensional can be characterized using a Bravais lattice as follows:
\begin{eqnarray*}
\RL_j=\{A_jn:n\in \mathbb{Z}^d\},\quad j=1,2,
\end{eqnarray*}
where $A_j\in \mathbb{R}^{d\times d}$ is an invertible matrix. 
The $j$-th layer unit cell can be represented as
\begin{eqnarray*}
\Gamma_j=\{A_j\alpha:\alpha\in[0,1)^d\},\quad j=1,2.
\end{eqnarray*}
Each layer $\RL_j ~(j=1,2)$ exhibits periodicity. This means that the layer remains invariant with respect to its lattice vectors
\begin{eqnarray*}
\RL_j = \RL_j + A_j n \qquad \forall ~n\in \mathbb{Z}^d.
\end{eqnarray*}
The corresponding reciprocal lattice and the reciprocal unit cell can be defined as:
\begin{eqnarray*}
\RL_j^*=\{2\pi A_j^{-T}n:n\in \mathbb{Z}^d\}
\qquad{\rm and}\qquad
\Gamma_j^*=\{2\pi A_j^{-T}\alpha:\alpha\in[0,1)^d\}
\end{eqnarray*}
respectively. 

Even though the layers $\RL_1$ and $\RL_2$ are periodic on their own, the translation invariance might not hold when the layers are stacked together. This leads to so-called incommensurate systems.

\begin{definition}
\label{incomm}
Two Bravais lattices $\RL_1$ and $\RL_2$ are called incommensurate if
\begin{eqnarray*}
\RL_1\cup\RL_2 + \tau = \RL_1\cup\RL_2 \quad \Leftrightarrow \quad \tau=\pmb{0}\in\R^d .
\end{eqnarray*}
\end{definition}

In this work, we take into account the systems such that not only the lattices $\RL_1$ and $\RL_2$ are incommensurate, but their corresponding reciprocal lattices $\RL^*_1$ and $\RL^*_2$ are also incommensurate. 
Our focus is primarily on the Schr\"{o}dinger operator for the bi-layer incommensurate system, which is given by
\begin{eqnarray}
\label{H}
H := -\frac{1}{2}\Delta + V_1 + V_2,
\end{eqnarray}
where $V_j:\R^d\rightarrow\R^{+}~(j=1,2)$ are smooth and $\RL_j$-periodic functions. 
Owing to the periodicity of the potentials, it is possible to express $v_j$ in terms of a Fourier series:
\begin{eqnarray*}
\label{V_series}
V_j(x) = \sum_{G\in \RL_j^*} \hat{V}_{j,G} e^{iG\cdot x} \quad{\rm with}\quad 
\hat{V}_{j,G} = \frac{1}{|\Gamma_j|}\int_{\Gamma_j}V_j(x)e^{-iG\cdot x}\dd x \qquad j=1,2.
\end{eqnarray*}

The Schr\"{o}dinger operator is a central object across numerous fields, including condensed matter physics and materials science.
It serves as a fundamental component in the mathematical modeling of quantum-physical processes, including Bose-Einstein condensates associated with ultracold bosonic or photonic gases and the electronic structure of molecule systems. 
In particular, we are interested in the incommensurate layered systems, such as single layers of low-dimensional materials stacked on top of each other with a twist. 
This configuration drastically affects the properties of the single layer counterparts, leading to the emergence of intriguing phenomena such as localization.

The localization is a fundamental phenomenon in both physics and materials science, characterized by the spatial confinement of particles within a specific region. This confinement results in a significant reduction of their mobility and propagation, thus deeply influencing various properties such as electrical conductivity, optical characteristics, and magnetism among others.
The implications of localization can be so substantial as to alter the inherent behavior of a material, potentially causing a transition from a metallic to an insulating state.
Consequently, the exploration of these mathematical manifestations not only assists in the quantitative understanding of localization effects but also offers a microscopic lens for comprehending the underlying physic.

By examining the Schr\"{o}dinger operator, one can derive various physical properties, including energy levels, electron density, and other measurable observables. 
Since the spectral behavior of the Schr\"{o}dinger operator proves to be especially fascinating with incommensurate potentials, the {\it density of states} (DoS) serves as an appropriate tool for examining its spectral distribution. This concept denotes the number of states within each energy interval at each energy level.
Another crucial concept in this field is the {\it integrated density of states} (IDoS), which is defined as the integral of the density of states from $-\infty$ to a real energy value $E$.
To effectively estimate the spectral distribution, Weyl's law offers an asymptotic formula for the spectral representation \cite{zworski2022semiclassical}:
\begin{equation}
    \label{weylslaw}
    N_V(E) : = (2\pi)^{-d}{\rm vol} \{(x, \xi) \in \Omega \times \R^d \ | \ V(x) + |\xi|^2 \leq E\} \quad {\rm as} \ E \to \infty,
\end{equation}
where $V:= V_1 + V_2$. In essence, this formula serves as a valuable tool for calculating the IDoS. It forms a significant connection between the spectral properties of an operator with the volume of certain subsets of phase space, opening up new avenues for exploration.
Furthermore, electron density describes the spatial probability distribution of electrons within the system, which fundamentally governs the chemical and physical properties of the systems and thus determines the macroscopic behavior of materials.
Specifically, the density operator for the incommensurate system is represented as:
\begin{equation*}
   \rho(x)=f_{\mu,\beta}(H)(x, x),
\end{equation*}
where $f_{\mu,\beta}(\cdot)=\frac{1}{1+e^{(\cdot-\mu)\beta}}$
is the Fermi-Dirac function, $\mu$ is the chemical potential and $\beta:=(k_{B}T)^{-1}$ is the inverse temperature with $k_B$ the Boltzmann constant.

\section{Effective potential of incommensurate systems}
\label{sec:effectiveP}

In a remarkable contribution, Filoche \& Mayboroda \cite{arnold2016effective} proposed a simple but astonishingly effective method to predict the structure of the eigenfunctions and the spectral distribution.
These predictions are attained by solving the corresponding partial differential source problem.
It makes it possible to predict the localization regions of electron states without the direct solution of eigenvalue problems, especially at a considerable computational cost. 

We present first the main features of the localization landscape theory introduced in \cite{arnold2019computing,arnold2016effective,filoche2012universal}. 
Following these ideas, 
the {\it landscape function} $u$ of incommensurate systems is defined as the unique solution to 
\begin{eqnarray}
   \label{IncommPDE}
    \Big(-\frac{1}{2}\Delta + V_1 + V_2\Big) u = 1.
\end{eqnarray}
In a bounded domain setting, one of the fascinating results is that the function $u$ can control point wisely all eigenfunctions of \eqref{H} by
\begin{equation}
\label{localeigendomain}
|\psi (x)| \leq \lambda u(x)\|\psi\|_{L^{\infty}}, \quad \forall \ x\in\Omega,
\end{equation}
where $\Omega \subset \R^d$ is an open, bounded domain. Equivalently, 
the eigenfunction $\psi$ can only localize in a bounded domain \cite{filoche2012universal}
\begin{equation*}
\{x: u(x)\geq 1/\lambda\}\subset \Omega.
\end{equation*}
By composing an eigenstate $\psi$ of $H$ as 
$\psi = u \tpsi$, we see that the auxiliary function $\tpsi$ obeys a Schr\"{o}dinger-type eigenvalue problem 
\begin{eqnarray*}
\label{effective_eigenproblem}
-\frac{1}{u^2} \frac{\partial}{\partial x}\Big(u^2 \frac{\partial}{\partial x} \tpsi\Big) + \frac{1}{u} \tpsi = \lambda\tpsi,
\end{eqnarray*}
in which the origin potential $V_1+V_2$ in \eqref{H} has disappeared.
Although the Laplacian operator is replaced by a slightly more complicated elliptic operator, the new function
\begin{eqnarray}
   \label{def:effectivepotential}
\veff(x) := u(x)^{-1}    
\end{eqnarray}
plays the role of effective potential and presents influential features. Numerical experiments \cite{ arnold2019computing,arnold2016effective, filoche2012universal} have suggested that the smallest local minima of $\veff$ correspond precisely to the location where the first few eigenfunctions localize. Moreover, the energy of the local fundamental state inside each well was found to be almost proportional to the value of the effective potential at its minimum inside the well \cite{arnold2019computing}.
Rather than predicting localization through \eqref{localeigendomain}, the effective potential \eqref{def:effectivepotential} incorporates abundant information, a superiority that has been theoretically and numerically substantiated in \cite{arnold2019localization,lemut2020localization,steinerberger2017localization,wang2020localization, wang2021exponential}.

Further, the effective potential serves as an approximation for the spectral distribution. In particular, by replacing the original potential with the effective potential, an approximation of the IDoS can be achieved. As outlined in the works of \cite{arnold2019computing, david2021landscape,filoche2012universal}, the IDoS, denoted as $N_{\rm eff}(E)$, can be represented as:
\begin{equation*}
\label{def:weyl_Neff}
   N_{\rm eff}(E) := (2\pi)^{-d} {\rm vol} \Big\{(x, \xi)\in \Omega \times \R^d ~ \big| ~ \veff(x) + |\xi|^2 \leq E \Big\}.
\end{equation*}
The variant, demonstrated through experimental results, exhibits high precision across the energy spectrum. Notably, it maintains remarkable accuracy even at lower energy levels, an accomplishment that is not typically realized with standard Weyl's law \eqref{weylslaw}.

\section{Plane wave discretizations}
\label{sec:pw}
\setcounter{equation}{0}

To present the numerical schemes for studying the localization of incommensurate systems, we first construct a discrete Hamiltonian by the plane wave basis. Let us reiterate the plane wave framework of incommensurate systems proposed in \cite{wang2022convergence, zhou2019plane} (c.f. also earlier work on quasicrystals \cite{jiang2014numerical}).

Let $W,~L >0$, we define the following truncation for the plane wave vectors in $\RL_1^* \times \RL_2^*$ introduced in \cite{wang2022convergence}
\begin{eqnarray*}
\label{set:cutoff}
\DD_{W,L} := \Big\{\big(G_{1},G_{2}\big) \in\RL_1^* \times \RL_2^* ~:~ \big|G_{1}+G_{2}\big|\leq W ,~\big|G_{1}-G_{2}\big|\leq L \Big\}.
\end{eqnarray*}
We consider the corresponding plane wave $\big\{e^{i(G_{1}+G_{2})\cdot x}\big\}_{(G_1, G_2)\in \DD_{W,L}}$ as basis functions, and they satisfies 
the following orthonormal condition:
\begin{eqnarray*}
\label{ortho}
\lim_{R\rightarrow\infty} \frac{1}{|B_R|}\int_{B_R} e^{-i(G_1+G_2)x} e^{i(G_1^{'}+G_2^{'})x} \dd x = \delta_{G_1G_1^{'}} \delta_{G_2G_2^{'}} \quad\forall~ (G_1, G_2), (G_1^{'}, G_2^{'}) 
\in \DD_{W,L},
\end{eqnarray*}
where $B_R\subset\R^d$ is the ball centered at the origin with radii $R$. With the plane wave discretization, we obtain the following discrete Hamiltonian of \eqref{H}
\begin{equation*}
\label{Hpw}
\hH^{W,L}_{\bG,\bG'} = \frac{1}{2}\big|G_1+G_2\big|^2\delta_{G_1G_1^{'}} \delta_{G_2G_2^{'}} + \hat{V}_{1,G_1-G_1^{'}} \delta_{G_2G_2^{'}} + \hat{V}_{2,G_2-G_2^{'}} \delta_{G_1G_1^{'}} 
\end{equation*}
for $(G_1,G_2),~(G_1^{'},G_2^{'}) \in \DD_{W,L}$.

Following that, we provide the process to predict the localization of the incommensurate system under plane wave discretizations.
With the plane wave basis functions, we can approximate the solution $u$ in \eqref{IncommPDE} by
\begin{eqnarray}
\label{uPW}
u(x) = \sum_{(G_1,G_2)\in \DD_{W,L}} \hu_{G_1,G_2} e^{i(G_1+G_2)\cdot x}.
\end{eqnarray}
Denote by 
$U:=\{\hu_{G_1,G_2}\}_{(G_1,G_2)\in \DD_{W, L}}$, we can derive the linear system by using the standard Galerkin projection and the orthonormal condition,
\begin{equation}
\label{pwlsystem}
 \hH^{W,L} U=I_{\bzero},
\end{equation}
where $\hH^{W,L} \in \C^{|\DD_{W,L}|\times |\DD_{W,L}|}$ and $I_{\bzero}:=\{\delta_{G_1\bzero}{\delta_{G_2\bzero}}\}_{(G_1,G_2)\in \DD_{W, L}}$.
Starting from the approximate solution $u$ \eqref{uPW}, we can compute directly the absolute value of the effective potential pointwise,
\begin{equation}
\label{pwVeff}
 |\veff(x)| = |u(x)^{-1}|, \qquad x \in \R^d.
\end{equation}
In a standard procedure of representing the spectral distribution and electron density under the plane wave discretizations, we first solve the matrix eigenvalue problem 
\begin{equation}
    \label{Heigen}
    \hH^{W,L} \Phi_j = \lambda_j \Phi_j,  \quad j = 1, \cdots, N_{W,L}.
\end{equation} 
By the definition of the integrated density of states, the eigenvalue counting function can be stated as 
\begin{equation*}
\label{def:IDos_countfunction}
    N(E) = \#\Big\{\lambda_j\leq E, ~ j= 1, \dots, N_{W,L} \Big\},
\end{equation*}
which represents the count of all eigenvalues less than or equal to some $E$.
Advancing further with the eigenvalue $\lambda_j$ and eigenvector $\Phi_j=\{\phi_{j,G_1,G_2}\}_{(G_1,G_2)\in \DD_{W, L}}$, we can express the discretized electron density immediately as:
\begin{equation}
\label{electronicdensity}
 \rho^{W,L}(x) = \frac{1}{S_{d,L}}\sum_{j=1}^{N_{W,L}}f_{\beta,\mu}(\lambda_j)|\psi_j(x)|^2,
\end{equation}
where $S_{d,L}$ is a normalized constant related to the dimension $d$ and parameter $L$ (see \cite{wang2022convergence} for more details), and $\psi_j(x)$ is approximated by the plane wave basis function, 
\begin{equation*}
\label{eigenfunction}
\psi_j(x) = \sum_{(G_1,G_2)\in \DD_{W,L}} \phi_{j,G_1,G_2} e^{i(G_1+G_2)\cdot x} \qquad {\rm for} \ x \in \R^d.
\end{equation*}
To predict the structure of the electron density, especially for identifying their localized domains, we first select a domain in real space and identify a series of local minima of $|\veff(x)|$. The location of these local minima will serve as our prediction of where the electron density localizes (as detailed in Section \ref{sec:comput}). The electron density not only circumvents the inaccuracy in predicting the order of localization positions for eigenstates when eigenvalues are close (as mentioned in \cite{arnold2019computing}), but also serves as a more fitting physical observable for studying localization in such systems.

Unlike the eigenvalue problem and related partial differential equation in \cite{arnold2019computing,filoche2012universal} and references therein, we do not first restrict our problem to a bounded domain equipped with a specific boundary condition at first, while directly discrete it by the plane wave basis function. 
The advantage is that we can study localization on the basis of our interests in the domain. 
We mention that this practice is not irresponsible, since these eigenvalue problems of incommensurate systems can be interpreted in a higher dimensional space with periodic boundary conditions (see \cite{zhou2019plane}).

\section{Numerical study}
\label{sec:comput}
\setcounter{equation}{0}

In this section, we will report some numerical experiments on the linear Schr\"{o}dinger operator for the incommensurate systems. Particularly, we compute the effective potentials under plane wave discretizations, analyze the localization of the electron density, and compute the spectral distribution. 
In order to assess the accuracy of the prediction approaches based on effective potential, the eigenvalues and eigenfunctions are computed by discretized eigenvalue problem \eqref{Heigen}.

{\bf Example 1.}
{\rm (1D incommensurate system)}
Considering the following eigenvalue problem: 
\begin{equation*}
    \label{exa1dH}
    \Big(-\frac{1}{2} \Delta +  V_1(x) + V_2(x) \Big) \psi(x) = \lambda \psi(x) \qquad x \in \R,
\end{equation*}
where 
\begin{equation}
\label{v_numerics}
V_1(x)= s_1\sum_{G_1\in \RL^*_1} e^{-\gamma|G_1|^2} e^{iG_1\cdot x} 
\qquad  {\rm and}  \qquad 
V_2(x)= s_2\sum_{G_2\in \RL^*_2} e^{-\gamma |G_2|^2} e^{iG_2 \cdot x}
\end{equation}
are incommensurate potentials with $s_1=3$, $s_2=2$, $L_1=2$, $L_2=\sqrt{5}-1$, and $\gamma = 0.05$.

\begin{figure}[!htb]
    \centering
    \includegraphics[height=8.0cm,width=14.0cm]{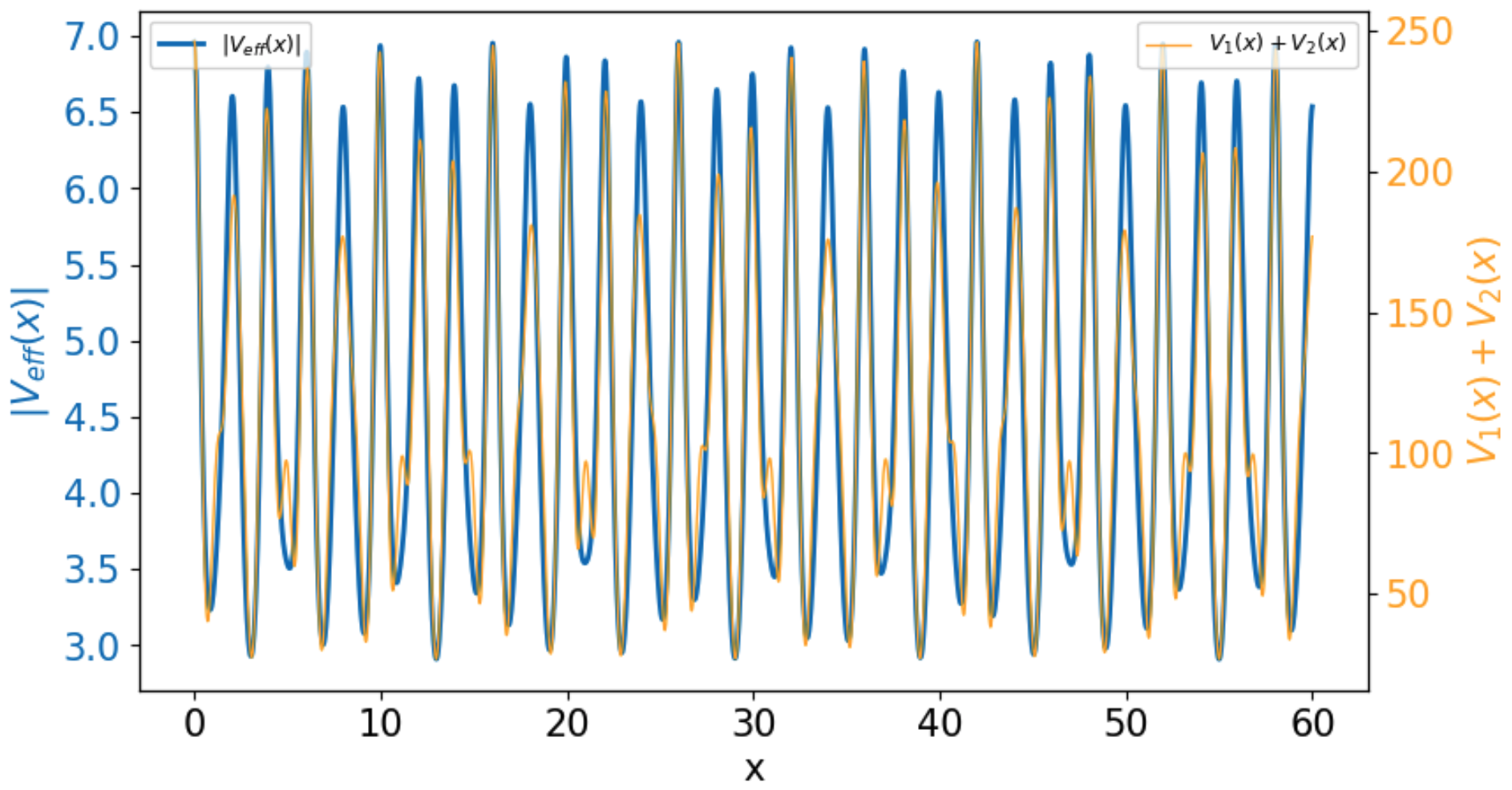}
    \caption{(Example 1) Effective potential and Incommensurate potential.}
    \label{fig:1D_Veff_V1V2}
\end{figure}

We tackle the linear systems as given in \eqref{pwlsystem} and compute the effective potential using \eqref{pwVeff}. The blue line in Fig. \ref{fig:1D_Veff_V1V2} depicts the effective potential $|V_{\rm eff}(x)|$. To contrast this with the original potential, we also display the incommensurate potential $V_1(x) + V_2(x)$ using a yellow line in the same figure. From this visualization, it is evident that both the effective potential and the original potentials capture the majority of the local minimum points. However, the effective potential, exhibiting a smoother profile than $V_1(x) + V_2(x)$, is able to refine a few local minima points.

\begin{figure}[!htb]
\centering
\begin{minipage}[t]{0.45\textwidth}
\centering
\includegraphics[height= 6.1cm]{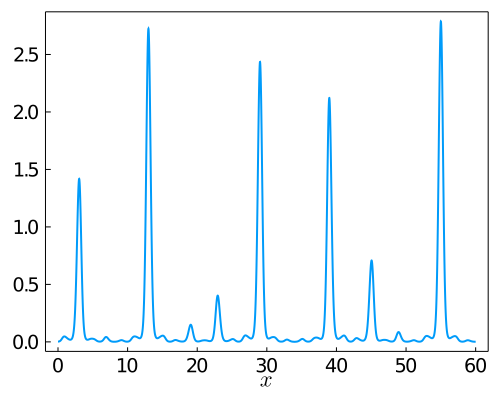}
\caption{(Example 1) Electron density $\rho^{W,L}(x)$ with $\mu = 3.8$.}
\label{fig:ex1:1Drhomu3p8}
\end{minipage}
\hskip 0.5cm
\begin{minipage}[t]{0.45\textwidth}
\centering
\includegraphics[height= 6.1cm]{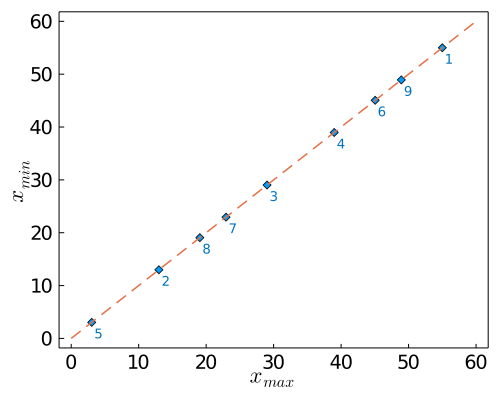}
\caption{(Example 1) The minima v.s maximum.}
\label{fig:ex1:minmax_1Drhomu3p8}
\end{minipage}
\end{figure}

\begin{figure}[!htb]
\centering
\begin{minipage}[t]{0.45\textwidth}
\centering
\includegraphics[height= 6.1cm]{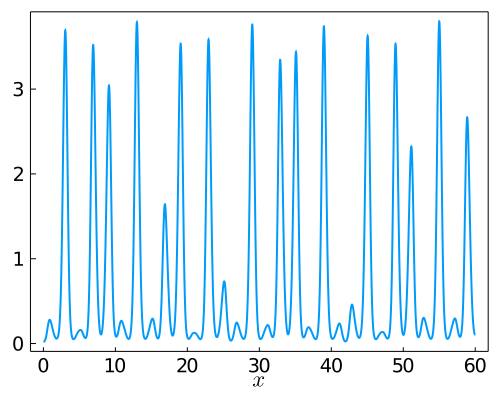}
\caption{(Example 1) Electron density $\rho^{W,L}(x)$ with $\mu = 4.0$.}
\label{fig:ex1:1Drhomu4}
\end{minipage}
\hskip 0.5cm
\begin{minipage}[t]{0.45\textwidth}
\centering
\includegraphics[height= 6.1cm]{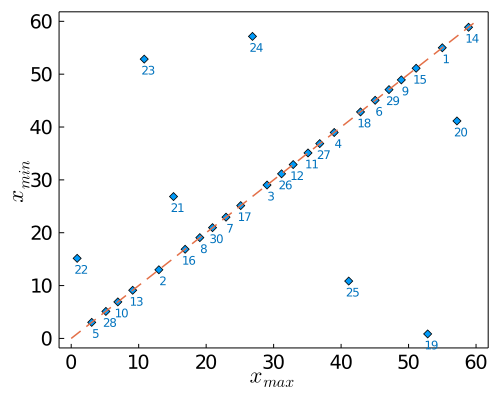}
\caption{(Example 1) The minima v.s maximum.}
\label{fig:ex1:minmax_1Drhomu4}
\end{minipage}
\end{figure}

\begin{figure}[!htb]
\centering
\begin{minipage}[t]{0.45\textwidth}
\centering
\includegraphics[height= 6.1cm]{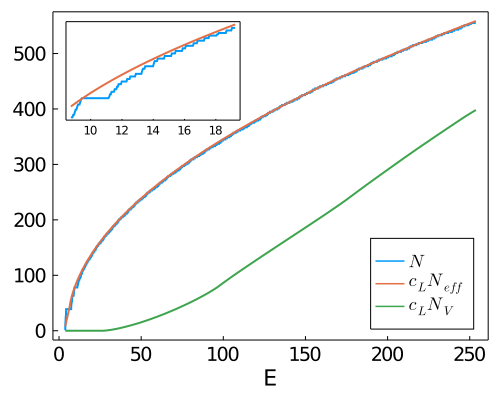}
\caption{(Example 1) Spectral distributions with $L = 200$ and $W = 50$.}
\label{fig:ex1:1dIDoS_L200W50}
\end{minipage}
\hskip 0.5cm
\begin{minipage}[t]{0.45\textwidth}
\centering
\includegraphics[height= 6.1cm]{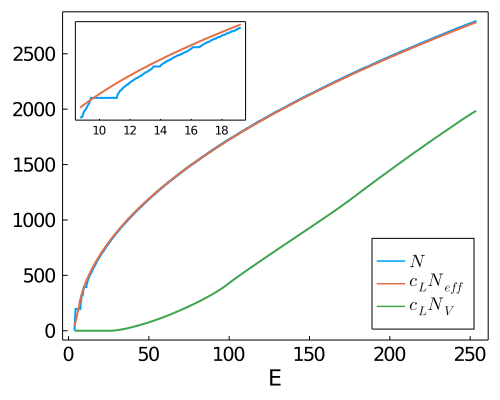}
\caption{(Example 1) Spectral distributions with $L = 1000$ and $W = 50$.}
\label{fig:ex1:1dIDoS_L1000W50}
\end{minipage}
\end{figure}

Further, we solve the eigenvalue problem as depicted in equation \eqref{Heigen} and proceed with computing the electronic density by \eqref{electronicdensity}. Specifically, our study revolves around the Fermi-Dirac function (see Section \ref{sec:Incommeig}) under the influence of varied parameters $\mu$ with an appropriate $\beta$. The parameter $\beta$ is associated with the inverse temperature of the system. In contrast,
the parameter $\mu$ is typically referred to as the chemical potential within the Fermi-Dirac distribution function. It is inherently associated with the system's electron density, effectively representing the number of electrons per unit volume.
This distribution outlines the probability of a state with a specified energy being occupied by an electron.
Consequently, adjusting the chemical potential allows for shifting the energy levels where electrons are most likely to be found, thus controlling the number of electrons within a specific energy range.
We compute $\rho^{W,L}(x)$ by taking different parameters $\mu$ with $W = 50, L = 1000$, $\beta = 100$.
The resultant electronic density is graphically represented in Fig. \ref{fig:ex1:1Drhomu3p8} and Fig. \ref{fig:ex1:1Drhomu4}, where it is evident that the electron density is localized within a certain region.
Furthermore, the local maxima of the electron density versus the local minima of the effective potential are shown in Figs. \ref{fig:ex1:minmax_1Drhomu3p8} and \ref{fig:ex1:minmax_1Drhomu4}, in which the numbers label the sequences of the minima and maxima. This comparison reveals that the regions of electron density localization coincide with the positions of the local minima of the effective potential except for a few points. 
The positions of the first $18$ local minima of the effective potential accurately correspond to the regions of electron density localization.
By comparing the electron density with the positions of local maxima and minima (Fig. \ref{fig:ex1:1Drhomu3p8} — Fig. \ref{fig:ex1:minmax_1Drhomu4}), we can infer that the localization of the electron density is more pronounced at the positions of smaller local minima.

On the other hand, to estimate the prediction of the spectral distribution, we employ different truncation parameters. Given the unique roles that parameters $L$ and $W$ play in modulating the spectral distribution (see \cite{wang2022convergence}), we factor this into our calculation by multiplying the prediction result by a constant related $L$. Intriguingly, as can be observed in Figs. \ref{fig:ex1:1dIDoS_L200W50} and \ref{fig:ex1:1dIDoS_L1000W50}, we find that the invariant Weyl's law, when amplified by a certain constant $c_L$, closely aligns with the eigenvalue count function. More specifically, the constant $c_L = 0.0185 L$ presented in the numerical experiments is determined through testing with various parameters $L$ and $W$. This adjustment enhances the precision of the approximation, taking into account the specific influences exerted by the size of the system. However, standard Weyl's law can not yield an accurate prediction for the incommensurate systems.

\begin{figure}[!htb]
  \centering
  \includegraphics[width=7.cm]{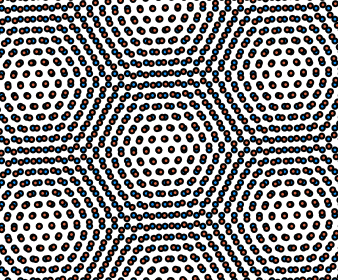}
  \caption{(Example 2) 2D incommensurate layered structures.}
   \label{fig:2D layers}
\end{figure}

{\bf Example 2} {\rm (2D incommensurate system)} 
Consider a two-dimensional incommensurate system that is created by overlaying two periodic lattices, where one layer is rotated by an angle $\theta=5^{\circ}$ relative to the other. 
More precisely, we take $\mathcal{R}_1= A_1\mathbb{Z}^2$ with 
\begin{equation*}
	A_1=2\left[ 
	\begin{gathered}
	\begin{matrix}
	1/2 &1/2\\-\sqrt{3}/2 & \sqrt{3}/2
	\end{matrix}
	\end{gathered}\right].	 
\end{equation*}
The structure of such twisted bi-layered systems is illustrated in Fig. \ref{fig:2D layers}. The potentials are characterized in the same manner as in \eqref{v_numerics} with $s_1 = 3, s_2 = 2, \gamma =0.05$.

\begin{figure}[!htb]
\centering
\begin{minipage}[t]{0.45\textwidth}
\centering
\includegraphics[height= 6.1cm]{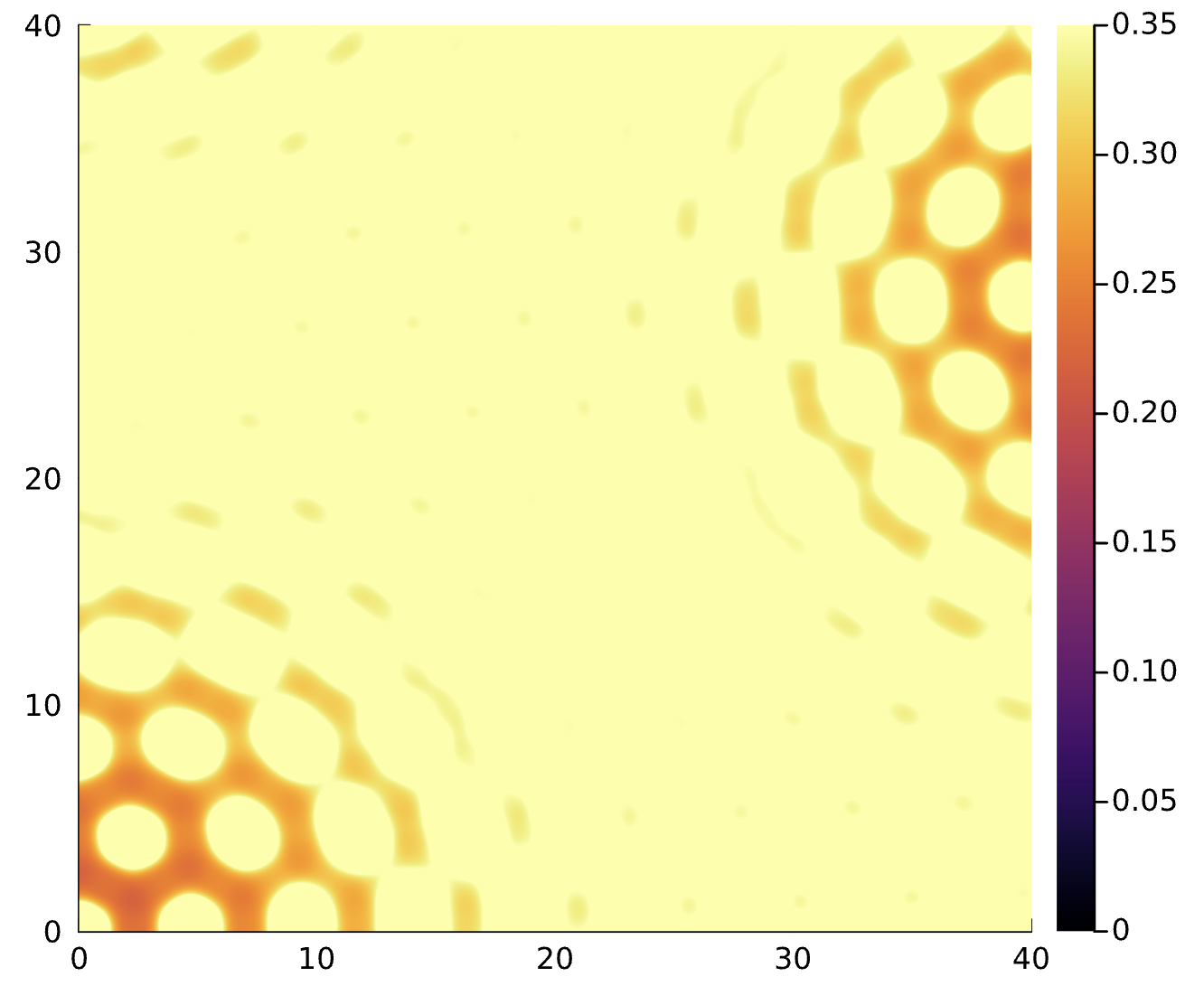}
\caption{(Example 2) Effective potential with rang $(0, 0.35)$.}
\label{fig:ex2:2dVeff_0to035}
\end{minipage}
\hskip 0.5cm
\begin{minipage}[t]{0.45\textwidth}
\centering
\includegraphics[height = 6.1cm]{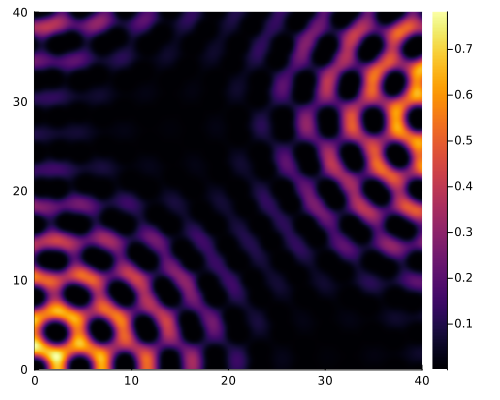}
\caption{(Example 2) Electron density $\rho^{W,L}(x)$ with $\mu = 0.5$.}
\label{fig:ex2:2drho50}
\end{minipage}
\end{figure}

\begin{figure}[!htb]
\centering
\begin{minipage}[t]{0.45\textwidth}
\centering
\includegraphics[height= 6.1cm]{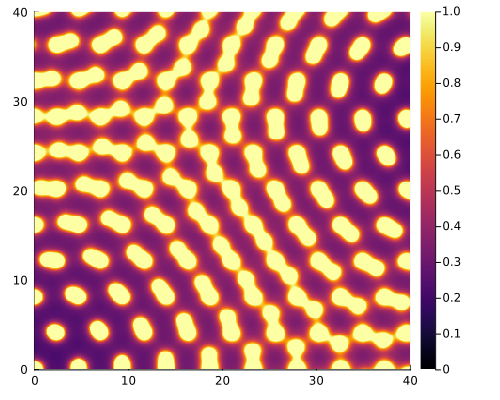}
\caption{(Example 2) Effective potential with range $(0, 1)$.}
\label{fig:ex2:2DVeff_L25W5}
\end{minipage}
\hskip 0.5cm
\begin{minipage}[t]{0.45\textwidth}
\centering
\includegraphics[height= 6.1cm]{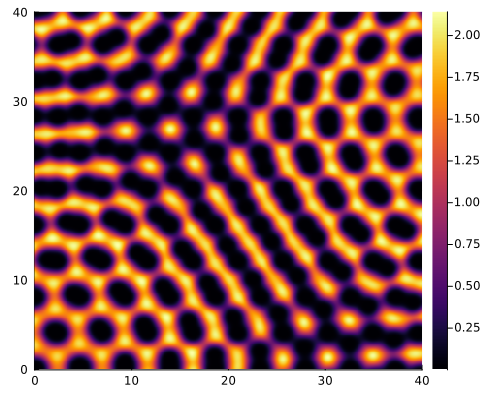}
\caption{(Example 2) Electron density 
$\rho^{W,L}(x)$ with $\mu = 1.0$.}
\label{fig:2Drho_mu1}
\end{minipage}
\end{figure}

\begin{figure}[!htb]
\centering
\begin{minipage}[t]{0.45\textwidth}
\centering
\includegraphics[height= 6.1cm]{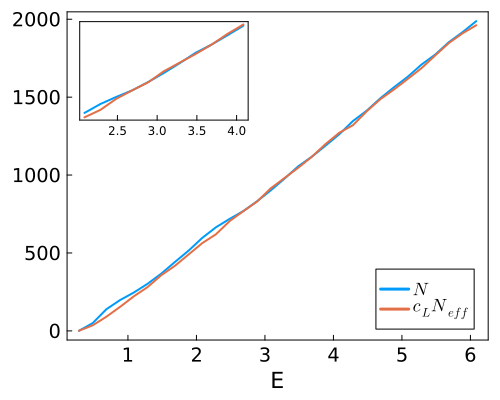}
\caption{(Example 2) Spectral distributions with $L = 20$ and $W = 5$.}
\label{fig:2D_IdosL20W5}
\end{minipage}
\hskip 0.6cm
\begin{minipage}[t]{0.45\textwidth}
\centering
\includegraphics[height= 6.1cm]{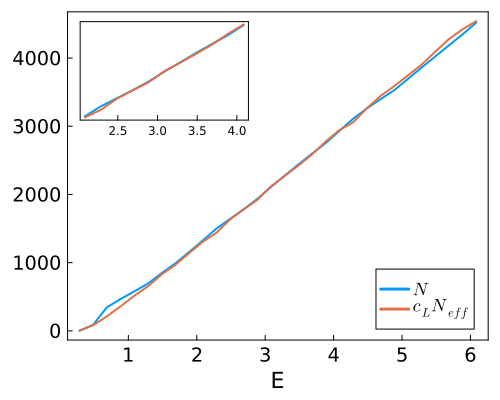}
\caption{(Example 2) Spectral distributions with $L = 30$ and $W = 5$.}
\label{fig:2D_IdosL30W5}
\end{minipage}
\end{figure}

We first plot the effective potential within a range of lower values. 
As depicted in Fig. \ref{fig:ex2:2dVeff_0to035} and Fig. \ref{fig:ex2:2drho50}, there is a remarkable correspondence between the minima of the effective potential and the positions of electron density localization. Expanding our examination to a broader range of the effective potential and increasing the chemical potential $\mu$, we observe intriguingly that the electron density tends to localize in the valleys of the effective potential. In other words, regions with larger values of electron density coincide perfectly with those where the effective potential values are smaller.

Similar to the one-dimensional case, we solve the eigenvalue problem \eqref{Heigen} by selecting different parameters $L$ and $W$. The eigenvalue counting function is then further compared with the invariant Weyl's law. From numerical results Fig. \ref{fig:2D_IdosL20W5} and Fig. \ref{fig:2D_IdosL30W5}, it is also demonstrated that the two are nearly consistent with each other by a constant factor related to $L$. Specifically, this constant factor is $c_L = 0.0072L^2$.
Integrating with the one-dimensional result, we can establish the dependence on the size of the reciprocal space and the dimensionality of the system.
The theoretical basis for this constant will be the subject of our subsequent research.

\section{Conclusion}
\label{sec:conclusion}

In this paper, we study the localization in the incommensurate systems under the plane wave discretizations with the aid of the effective potential. Our study uniquely ventures into exploring these problems from the perspective of the entire real space.
With the plane wave presentation of the effective potential, we numerically exhibit that electron density, 
localizes in regions where the effective potential exhibits local minima in good agreement with direct eigenpairs calculations and intuitive.
We further provide a robust prediction for the spectral distribution from the variant of Weyl's law utilizing the effective potentials.
Our future works will involve applications in the more practical incommensurate systems and study the deeper connection between the effective potentials and other physical quantities of interests.

\vskip 0.3cm

\noindent
{\bf Acknowledgements.}
The authors gratefully thank Prof. Huajie Chen and Dr. Daniel Massatt for their insightful discussions. This work was supported by the National Key R \& D Program of China under grants 2019YFA0709600 and 2019YFA0709601. Y. Zhou’s work was also partially supported by the National Natural Science Foundation of China under grant 12004047.

\small
\bibliographystyle{abbrv}
\bibliography{locality_incomm}

\end{document}